\documentclass[preprint,aps]{revtex4-1}
\usepackage[british]{babel}
\usepackage{amsmath,graphicx,alltt}

\newcommand{\mathLaplace}{\Delta}
\newcommand{\tmmathbf}[1]{\ensuremath{\boldsymbol{#1}}}
\newcommand{\tmop}[1]{\ensuremath{\operatorname{#1}}}

\newcommand{\tmtextit}[1]{\text{{\itshape{#1}}}}
\newcommand{\tmverbatim}[1]{\text{{\ttfamily{#1}}}}

\begin{document}

\title{The theory of electromagnetic line waves}

\author{S. A. R. Horsley}
\email[Email: ]{s.horsley@exeter.ac.uk}
\author{A. Dwivedi}
\affiliation{School of Physics and Astronomy,\\
University of Exeter,\\
Stocker Road,\\
EX4 4QL}

\

\begin{abstract}
  Whereas electromagnetic \tmtextit{surface waves} are confined to a planar interface between \tmtextit{two}
  media, \tmtextit{line waves} exist at the one--dimensional interface between \tmtextit{three} materials.  Here we derive a non--local integral equation for computing the properties of line waves, valid for surfaces characterised in terms of a general tensorial impedance.  We find a good approximation---in many cases---is to approximate this as a local differential equation, where line waves are one--dimensional analogues of surface plasmons bound to a spatially dispersive metal.  For anisotropic surfaces we find the oscillating decay of recently discovered `ghost' line waves can be explained in terms of an effective gauge field induced by the surface anisotropy.  These findings are validated using finite element simulations.
\end{abstract}

{\maketitle}

\section{Introduction}

Between appropriate materials, most waves can become a surface wave: a mode
trapped at a planar interface between different media.  Electromagnetic
(EM) surface waves between bulk media include surface plasmons,
magnetoplasmons, and Dyakonov waves, which depend on bulk material
properties.  Meanwhile, Tamm states can be trapped between periodically
layered media, without the need for negative permittivity, permeability, or
anisotropy.  For a review of this rich EM surface wave taxonomy see Ref.~\cite{polo2013}.  Metasurfaces can be used to control these surface waves, providing a structured interface that imposes an effective boundary
condition, either confining the wave or modifying its propagation characteristics~\cite{martini2015,sievenpiper2018}.  Similarly, at an interface between elastic materials there are Rayleigh and Love surface waves~\cite{volume7}, which can be controlled using e.g. seismic
metamaterials~\cite{brule2014}.  The existence of both elastic and EM surface waves can be inferred from topological arguments~\cite{bliokh2019,bliokh2019b}, which additionally predicts the appearance of unidirectional surface waves between a wide range of gyrotropic and bianisotropic materials~\cite{khanikaev2013,horsley2019,horsley2021}.

\begin{figure}[h]
  \raisebox{0.0\height}{\includegraphics[width=6.65656565656566cm,height=7.21418732782369cm]{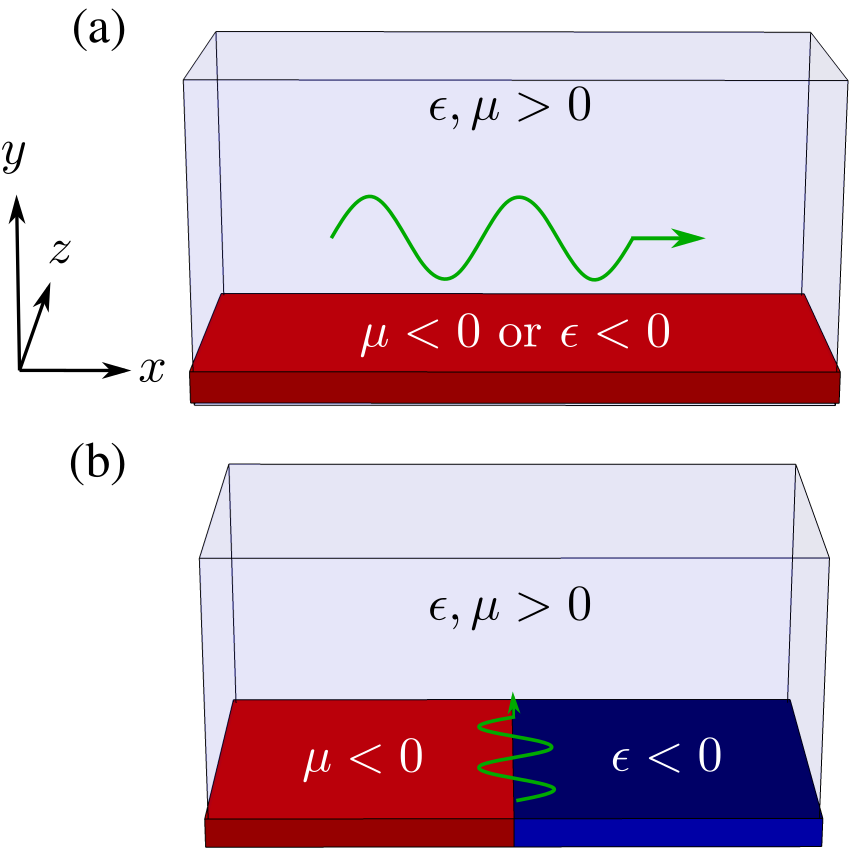}}
  \caption{\tmtextit{Surface versus line waves:} (a) Two dimensional
  \tmtextit{surface waves} (plasmons or magnetoplasmons) are bound to the
  interface between an ordinary dielectric (e.g. air or glass) and a material
  with either negative permittiviy, $\varepsilon$ or permeability, $\mu$. (b)
  One dimensional \tmtextit{line waves} are bound to the interface between
  \tmtextit{three} materials; an ordinary dielectric ($y > 0$), a negative
  permittivity material ($y < 0, x > 0$), and a negative permeability material
  ($y < 0, x < 0$).\label{fig:line-mode-schematic}}
\end{figure}

Despite a wealth of previous work on these two dimensional \tmtextit{surface} waves, their one dimensional cousins, `\tmtextit{line waves}' are much less well understood. A schematic example is shown in Fig. \ref{fig:line-mode-schematic}b, where a line wave occurs at the common, one dimensional interface where either the permittivity, permeability, or both change sign.  This new type of excitation was predicted and experimentally confirmed in Refs.~\cite{horsley2014} and~\cite{sievenpiper2017}, and further work has
examined their relationship to topology \cite{sievenpiper2019}, spin--momentum locking~\cite{xu2022}, and found instances of line waves between non--Hermitian parity--time symmetric media~\cite{moccia2020}.

To--date there have been three approaches to the theory of line waves: (i) initially \cite{horsley2014} the analysis was restricted to the asymptotic (electrostatic and magnetostatic) limit, where the propagation
constant $k$ takes a large value $k \gg \omega / c$, finding that the mode requires the surface impedance of the lower two media (see Fig. \ref{fig:line-mode-schematic}b) to be equal and opposite.  This result is
analogous to the asymptotic limit of the surface plasmon/magnon, where the bounding media have equal and opposite values of the permittivity/permeability.  Although this asymptotic result agrees with numerical simulations, it gives no indication of the dispersion of the mode or the general conditions for its existence.  Meanwhile, (ii) in Ref. \cite{sievenpiper2017} the authors found an exact analytical solution, a
solution restricted to the case of lower media that are perfect electric and magnetic conductors, an unusual case where the propagation constant becomes independent of frequency.  Finally, (iii) in Ref.~\cite{kong2019} the authors used Sommerfeld--Malyuzhinets diffraction theory to find a general exact solution.  Although exact, it requires the numerical evaluation of the Malyuzhinets function (expressed as an exponential of a numerical
integral) and to determine e.g. the propagation constant we must numerically find a combination of these functions that vanishes.  So far it has been challenging to generalize this exact solution to other types of wave where the Malyuzhinets solution does not apply (e.g. elastic waves, or more general types of electromagnetic media), and as a result most authors resort to finite element simulations~\cite{moccia2023}.

The aim here is to derive a simple, approximate, yet accurate theory of line waves, allowing us to build an intuitive theory that can be generalized to other kinds of bounding media and wave types.  The difficulty of deriving this theory compared to, say the characteristics of the surface plasmon, lies in the difference in the dimensionality of the wave propagation and three dimensional space. For a surface wave both the frequency, $\omega$ and the two component in--plane wave--vector, $\tmmathbf{k}_{| |}$ are conserved, meaning that the remaining decay constant away from the surface can be written entirely in terms of these \tmtextit{three} conserved quantities via the dispersion relation. All that remains of the problem is to then find the values of $\omega$ and $\tmmathbf{k}_{| |}$ such that the boundary conditions are satisfied.

By contrast a line wave only has \tmtextit{two} conserved quantities, the frequency and the wave--vector component $k_z$, directed along the common interface (see Fig. \ref{fig:line-mode-schematic}b).  These conserved
quantities do not provide enough information to determine the form of the field in e.g. the $x$--$y$ plane and we must therefore find \tmtextit{both} the field distribution and the values of the conserved quantities such that the boundary conditions are satisfied.

Here we sidestep this difficulty through writing an effective wave equation for the field on the \tmtextit{surface alone}.  Interestingly this equation is very similar to the two dimensional vector Helmholtz equation that would hold for the surface magnetic field $\tmmathbf{H}_{| |}$ in strictly two dimensional space, with the surface impedance playing the role of the permeability.  The third dimension is encoded in a non--local integral kernel, which serves to `blur' this two--dimensional equation.  As we shall show, it is simple to apply and approximate this effective surface wave equation to analyse the properties of line waves and understand new results such as the oscillatory decay of `ghost' line waves {\cite{moccia2023}}.

\section{The Non--local surface wave equation}

Rather than consider the bulk media sketched in
Fig.\ref{fig:line-mode-schematic}, where we would have to consider the wave in
the region $y < 0$, it is simpler to characterize the materials in the half
space $y < 0$ in terms of a boundary condtion at $y = 0$: in terms of a
surface impedance.  To achieve this on the surface $y = 0$, we take the
electromagnetic field to satisfy an impedance boundary condition~\cite{sievenpiperthesis,senior1995},
\begin{equation}
  \tmmathbf{E}_{| |} = \text{i} \eta_0 \chi (x)  \tmmathbf{\hat{y} \times
  H}_{| |} \label{eq:impedance-bc}
\end{equation}
where $\chi (x)$ is the surface reactance (surface impedance, $Z = \sqrt{\mu /
\varepsilon} = \text{i} \chi$).\quad Our first aim is to re--write this
equation as an effective wave equation, written entirely in terms of the
behaviour of the in--plane $\tmmathbf{H}$ field on the $y = 0$ surface.  To eliminate the in--plane electric field field from (\ref{eq:impedance-bc})
we apply the Maxwell equation, $\tmmathbf{\nabla \times} \eta_0 \tmmathbf{H} =
- {\rm i} k_0 \tmmathbf{E}$,
\begin{equation}
  \tmmathbf{E}_{| |} = \frac{\text{i} \eta_0}{k_0}  \tmmathbf{\hat{y} \times}
  \left( \frac{\partial \tmmathbf{H}_{| |}}{\partial y} - \tmmathbf{\nabla}_{|
  |} H_y \tmmathbf{} \right) \label{eq:in-plane-field}
\end{equation}
where $k_0 = \omega / c$ and $\eta_0 = \sqrt{\mu_0 / \varepsilon_0}$.

Although Eqns. (\ref{eq:impedance-bc}--\ref{eq:in-plane-field}) can be
combined into a single equation that describes the components of the magnetic
field on the surface alone, it is not yet the `surface wave equation' we want:
the equation does not make reference to the behaviour of the field in the
plane of the surface alone.  We therefore cannot yet find a solution using
data on the surface alone until we have removed the third dimension entirely,
eliminating both the unknown derivatives normal to the plane, and the normal
field component $H_y$.

The out of plane field components and derivatives can be eliminated from Eq.
(\ref{eq:in-plane-field}) through using a Fourier decomposition of the
magnetic field, in terms of Fourier amplitudes $\tilde{\tmmathbf{H}}_{|
|}$.\quad For instance, $\partial \tmmathbf{H}_{| |} / \partial y$ can be
written as an in--plane convolution of $\tmmathbf{H}_{| |}$ with a kernel $K
(x - x'),$
\begin{align}
    \left. \frac{\partial \tmmathbf{H}_{| |}}{\partial y} \right|_{y = 0}  &=
    - \text{e}^{\text{i} k_z z}  \int_{- \infty}^{\infty} \frac{\text{d} k}{2
    \pi}  \sqrt{k^2 + \kappa^2} \tmmathbf{\tilde{H}}_{| |} (k) 
    \text{e}^{\text{i} k x}\nonumber\\
    &= \int_{- \infty}^{\infty} \text{d} x' K (x - x')  \left(
    {\tmmathbf{\nabla}_{| |}'}^2 + k_0^2 \right)  \tmmathbf{H}_{| |} (x', z) \label{eq:kernel-1}
\end{align}
where $\kappa = (k_z^2 - k_0^2)^{1 / 2}$.\quad The integration kernel in Eq.
(\ref{eq:kernel-1}) is proportional to a modified Bessel function of zeroth
order~\cite{NIST:DLMF}
\begin{align}
    K (x - x') & = \int_{- \infty}^{\infty} \frac{\text{d} \xi}{2 \pi}
    \frac{\text{e}^{\text{i} \xi \kappa (x - x')}}{\sqrt{\xi^2 + 1}}\nonumber\\
    & = \frac{1}{\pi} K_0 (\kappa | x - x' |).\label{eq:kernel-defn}
\end{align}
In the above we assume $k_z > k_0$ so that---via momentum conservation---the
surface wave remains confined, whatever the inhomogeneity of the surface
impedance in $x$.\quad This assumption leads to the real valued modified
Bessel function in (\ref{eq:kernel-defn}).\quad For $k_z < k_0$, the kernel
becomes complex valued, ultimately making our problem non--Hermitian due to
the scattering of the wave into the space above the surface.

\begin{figure}[h]
  \raisebox{0.0\height}{\includegraphics[width=8.05060343696707cm,height=6.02923717696445cm]{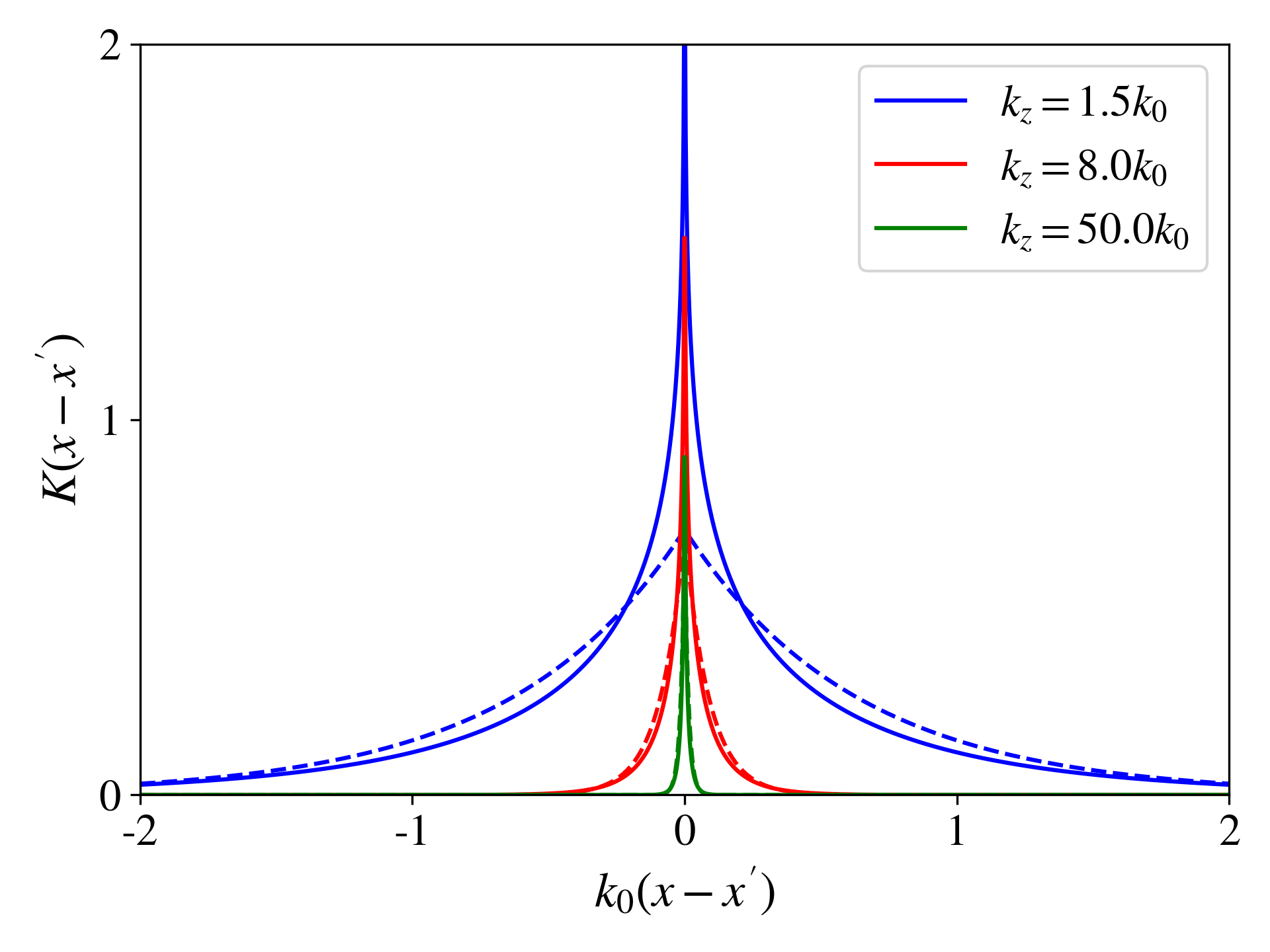}}
  \caption{\tmtextit{Kernel in the non--local Helmholtz equation}
  (\ref{eq:surface-eqn}): Defined in Eq. (\ref{eq:kernel-defn}), the integral
  kernel appearing in the surface Helmholtz equation (\ref{eq:surface-eqn})
  becomes increasingly localized as the propagation constant $k_z / k_0$ is
  increased.  Dashed lines show the approximation $K^{(1)}$ given in
  (\ref{eq:K1}), which becomes increasingly accurate as $k_z / k_0$ is
  increased.\label{fig:kernel}}
\end{figure}

The basic idea of this section is encapsulated in Eq. (\ref{eq:kernel-1}): we
can reformulate the boundary condition (\ref{eq:impedance-bc}) entirely in
terms of the behaviour of the magnetic field on the surface alone. The price is that the resulting equation is non--local, involving an exponentially localized kernel $K (x - x')$ that in Eq. (\ref{eq:kernel-1}) averages the Helmholtz equation over the surface (for the form of the kernel see Fig. \ref{fig:kernel}) .  The same calculation can also be performed to eliminate the out of plane
magnetic field $H_y$ from (\ref{eq:in-plane-field}), using the condition
$\tmmathbf{\nabla \cdot H} = 0$
\begin{equation}
  H_y = \int \text{d} x'  \tmmathbf{\nabla}_{| |}' \cdot \tmmathbf{H}_{| |}
  (x') K (x - x') \label{eq:kernel-2}
\end{equation}
Substituting Eqns. (\ref{eq:kernel-1}) and (\ref{eq:kernel-2}) into
(\ref{eq:impedance-bc}) and (\ref{eq:in-plane-field}) then gives us the final
non--local equation governing the field on the surface
\begin{equation}
  \int_{- \infty}^{\infty} \text{d} x' K (x - x')  (\tmmathbf{\nabla}_{| |}'
  \times \tmmathbf{\nabla}_{| |}' \times - k_0^2)  \tmmathbf{H}_{| |} (x')
  +k_0 \chi (x)  \tmmathbf{H}_{| |} = 0 \label{eq:surface-eqn}
\end{equation}
This is the equation that governs the field on an impedance boundary, making
no reference to the space above the surface. To some, Eq.
(\ref{eq:surface-eqn}) might be a pleasing result: it is a generalization of
the two dimensional vector Helmholtz equation $\tmmathbf{\nabla}_{| |} \times
\tmmathbf{\nabla}_{| |} \times \tmmathbf{H}_{| |} - k_0^2 \mu \tmmathbf{H}_{|
|} = 0$, the free space equation appearing within the integrand of
(\ref{eq:surface-eqn}), and the surface reactance playing a role similar to a
magnetic susceptibility.  The fact that there is a third dimension normal
to the surface is encoded in the integral kernel $K (x - x')$ defined in Eq.
(\ref{eq:kernel-defn}), which acts to `blur' the wave operator
$\tmmathbf{\nabla}_{| |} \times \tmmathbf{\nabla}_{| |} \times - k_0^2$ on the
surface.  As shown in Fig. \ref{fig:kernel}, as the propagation constant
$k_z$ is increased, this blurring reduces, reflecting the increasing
confinement of the field to the surface.

\subsection{Example: propagation on a homogeneous surface}

Before applying our integral equation (\ref{eq:surface-eqn}) to line waves,
the reader might need convincing that this equation reproduces known
results.  For a surface with a uniform impedance it is simple to solve Eq.
(\ref{eq:surface-eqn}).  Given it's Fourier representation
(\ref{eq:kernel-defn}), the integral kernel $K (x - x')$ has plane wave
eigenfunctions,
\begin{equation}
  \int_{- \infty}^{\infty} \text{d} x' K (x - x') \text{e}^{\text{i} k x'} =
  \frac{\text{e}^{\text{i} k x}}{\sqrt{k^2 + \kappa^2}}
  \label{eq:kernel-eigenfunction}
\end{equation}
Therefore, writing the in--plane magnetic field as a plane wave times a
constant vector amplitude, $\tmmathbf{H}_{| |} (x) = \tmmathbf{H}_0\, 
\text{e}^{\text{i} k x}$, the integral equation (\ref{eq:surface-eqn}) becomes
a simple algebraic equation that is identical to the dispersion relation for
electromagnetic waves in a magnetic material in two dimensions
\begin{equation}
  \left[ \tmmathbf{k}  \times \tmmathbf{k}  \times + k_0^2 \;
  \mu_{\tmop{eff}} (\tmmathbf{k}) \right]  \tmmathbf{H}_0 = 0
  \label{eq:uniform-surface}
\end{equation}
where $\tmmathbf{k} = k \tmmathbf{\hat{x}} + k_z \tmmathbf{\hat{z}}$.  The
effective permeability in Eq. (\ref{eq:uniform-surface}) is given by
\begin{equation}
  \mu_{\tmop{eff}} (\tmmathbf{k}) = 1 - \frac{\sqrt{k^2 + \kappa^2}}{k_0} \chi
  . \label{eq:mueff}
\end{equation}
Interestingly, unlike the ordinary dispersion relation for electromagnetic
waves in two dimensions, the effective permeability for our surface waves
(\ref{eq:mueff}) depends on the in--plane wave--vector $\tmmathbf{k}$.
 Surface waves bound to a uniform impedance boundary are thus equivalent to
two--dimensional electromagnetic waves in a spatially dispersive magnetic
material.  There is a very good physical reason for this: if the effective
permittivity (\ref{eq:mueff}) was \tmtextit{not} spatially dispersive, there
would only be a single transverse magnetic (TM) surface mode with dispersion
relation $k^2 + k_z^2 = \mu_{\tmop{eff}} k_0^2$ (there is only one transverse
polarization in two dimensions).  The fact that the effective permeability
(\ref{eq:mueff}) is $\tmmathbf{k}$ dependent means that there are both
longitudinal (TE) and transverse (TM) wave solutions to the two dimensional
equation (\ref{eq:uniform-surface}).

The longitudinal (TE) surface mode can be derived through performing an inner
product of (\ref{eq:uniform-surface}) with $\tmmathbf{k}$, which reduces the
dispersion relation to the condition $\mu_{\tmop{eff}} (\tmmathbf{k}) = 0$,
familiar from the theory of longitudinal modes in spatially dispersive
crystals {\cite{hopfield1963}},
\begin{equation}
  \; \mu_{\tmop{eff}} (\tmmathbf{k})  (\tmmathbf{k}_{| |} \cdot
  \tmmathbf{H}_0) = \left[ k_0 - \sqrt{k^2 + \kappa^2} \chi \right] 
  (\tmmathbf{k}_{| |} \cdot \tmmathbf{H}_0) = 0 \label{eq:longitudinal-mode}
\end{equation}
i.e. $(k^2 + \kappa^2)^{1 / 2} = k_0 / \chi$.  This condition can only be
fulfilled when the surface reactance is \tmtextit{positive}.  Meanwhile,
taking $\tmmathbf{k \cdot H}_{| |} = 0$ in (\ref{eq:uniform-surface}) yields
the dispersion relation for transverse (TM) surface modes,
\begin{equation}
  \left[ (k^2 + \kappa^2) + k_0 \sqrt{k^2 + \kappa^2} \chi \right]
  \tmmathbf{H}_0 = 0 \label{eq:transverse-mode}
\end{equation}
i.e. $(k^2 + \kappa^2)^{1 / 2} = - k_0 \chi$, which can only be fulfilled when
the surface reactance is \tmtextit{negative}. Equations
(\ref{eq:longitudinal-mode}) and (\ref{eq:transverse-mode}) are the
well--known dispersion relations for transverse electric (TE) and transverse
magnetic (TM) surface waves \cite{sievenpiperthesis,tretyakov2003},
here derived as a special case of the non--local surface Helmholtz equation
(\ref{eq:surface-eqn}).

\section{Line Waves on isotropic surfaces}\label{sec:line-wave}

We now apply the non--local surface wave equation (\ref{eq:surface-eqn}) to
the central problem of this paper: electromagnetic line waves confined to the
$x = 0$ interface between two impedance boundaries.  To set up this
problem we write the surface reactance distribution representing the materials
sketched in Fig. \ref{fig:line-mode-schematic}b in terms of the average
reactance $\chi_b$, and the contrast $\Delta \chi$,
\begin{equation}
  \chi (x) = \chi_b + \frac{\Delta \chi}{2} \tmop{sign} (x)
  \label{eq:reactance-line-wave} .
\end{equation}
With this identification, the surface integral equation (\ref{eq:surface-eqn})
becomes an eigenvalue problem for the average reactance $\chi_b$
\begin{equation}
  - \frac{1}{k_0} \int_{- \infty}^{\infty} \text{d} x' K (x - x') 
  (\tmmathbf{\nabla}_{| |}' \times \tmmathbf{\nabla}_{| |}' \times - k_0^2) 
  \tmmathbf{H}_{| |} (x')
  - \frac{1}{2} \Delta \chi \tmop{sign} (x) 
  \tmmathbf{H}_{| |} = \chi_b^{(0)}  \tmmathbf{H}_{| |} .
  \label{eq:eigenvalue-problem}
\end{equation}
The superscript `$(0)$' has been added to the average reactance to indicate
that this is the eigenvalue of the equation without approximations.  We
thus determine the line mode dispersion relation through specifying the
wave--number $k_0$, the $k_z$ wave--vector component, and the reactance contrast
$\Delta \chi$.  The eigenvalue of the integral equation
(\ref{eq:eigenvalue-problem}) then gives the average surface reactance
required such that the given mode $\tmmathbf{H}_{| |}$ is supported for these
values of $k_0$, $k_z$, and $\Delta \chi$.

Numerically it is simplest to solve (\ref{eq:eigenvalue-problem}) in the
Fourier domain where using the eigenfunctions of the kernel
(\ref{eq:kernel-eigenfunction}) the integral operator becomes a simple
multiplication, and the term proportional to $\Delta \chi$ becomes a principal
value integral
\begin{equation}
  \frac{1}{k_0}  \left( \tmmathbf{} \frac{\tmmathbf{k} \times \tmmathbf{k}
  \times + k_0^2}{\sqrt{k^2 + \kappa^2}} \right) \cdot
  \tilde{\tmmathbf{H}}_{| |} (k)
  + \frac{\text{i} \Delta \chi}{2} 
  \frac{1}{\pi} \text{P}  \int_{- \infty}^{\infty}
  \frac{\tilde{\tmmathbf{H}}_{| |} (k')}{k - k'} \text{d} k' =
  \chi_b^{(0)}  \tilde{\tmmathbf{H}}_{| |} (k),
  \label{eq:fourier-eigenvalue}
\end{equation}
where $\tilde{\tmmathbf{H}}_{| |}$ is the Fourier transform of the
in--plane magnetic field, and `P' indicates the principal part of the
integral, which itself is a Hilbert transform~\cite{king2009}.\quad The
appearance of a Hilbert transform is expected in this problem: it's
eigenfunctions are the functions of $k$ that are analytic in either the upper
or lower half of the complex $k$ plane, which represent real space functions
that are confined to either side of the $x = 0$ line, where the reactance
takes a fixed value.

Equation (\ref{eq:fourier-eigenvalue}) can be solved using the Wiener--Hopf
method~\cite{kisil2021}, although we don't pursue this here.  It is
also straightforward to solve this equation numerically.  To do this we discretize both the $x$--axis and $k$--space into $N$ points and write the left hand side of Eq. (\ref{eq:fourier-eigenvalue}) as a $2 N \times 2 N$ matrix with the Hilbert transform written as $\hat{\mathcal{H}} = - \text{i} \hat{\mathcal{F}}^{-
1} \tmop{diag} (\tmop{sign} (x_n)) \hat{\mathcal{F}}$ implemented in terms
of the discrete Fourier transform matrix $\hat{\mathcal{F}}_{n m} = \exp
\left( - \text{i} k_n x_m \right)$.  When written in this matrix form it
is evident that, for real $\mathLaplace \chi$ the left hand side of
(\ref{eq:fourier-eigenvalue}) is a Hermitian operator with corresponding real
eigenvalues, $\chi_b^{(0)}$.  For purely imaginary $\Delta \chi$ the
system is PT--symmetric, and the operator on the left of
(\ref{eq:fourier-eigenvalue}) is equivalent to a real--valued but
non--Hermitian matrix with eigenvalues that are thus real, or in complex
conjugate pairs.  This is consistent with the findings of~\cite{moccia2020}, where it was found that line waves can be supported on
impedance surfaces with balanced loss and gain.

\begin{figure}[h]
  \raisebox{0.0\height}{\includegraphics[width=15.9966548602912cm,height=5.33221828676374cm]{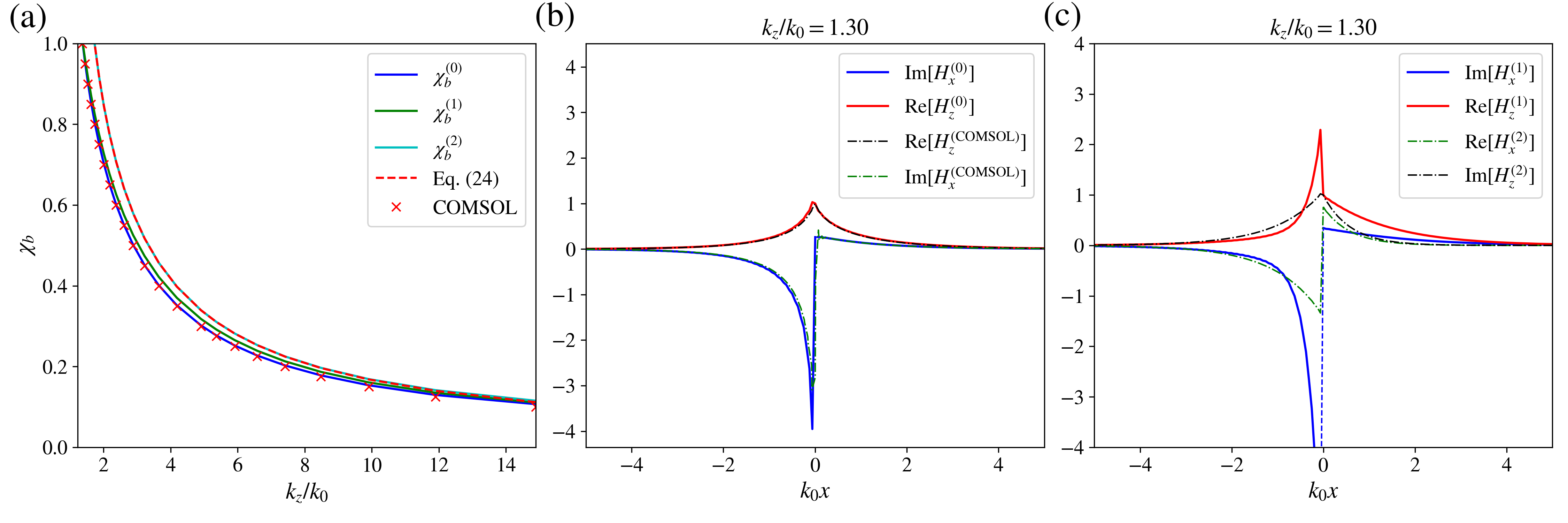}}
  \caption{\tmtextit{Line mode dispersion and field profiles:} (a) Dispersion
  relation ($\chi_b$ as a function of $k_z$, for a fixed reactance contrast
  $\Delta \chi = 2.309$) numerically calculated using exact integral equation
  (\ref{eq:fourier-eigenvalue}), the two approximations given in (\ref{eq:K1})
  and (\ref{eq:K2}), and analytic relation
  (\ref{eq:analytic-dispersion}).\quad The $800 \times 800$ discrete Fourier
  transform matrix was constructed using spatial periodicity $L = 8 \lambda$.
  \quad Red crosses show calculated values of $k_z / k_0$ computed using the
  COMSOL Multiphysics eigenvalue solver.\quad Note that the results of Eqns.
  (\ref{eq:K2}) (cyan) and (\ref{eq:analytic-dispersion}) (red dashed) are
  identical.\quad (b) Comparison of in--plane magnetic field profiles
  $\tmmathbf{H}_{| |}$ calculated from the exact integral equation
  (\ref{eq:fourier-eigenvalue}) and from COMSOL Multiphysics. (c) Field
  profiles calculated using approximations (\ref{eq:K1}) and
  (\ref{eq:K2}).\label{fig:dispersion}}
\end{figure}

As the eigenvectors of the operator (\ref{eq:fourier-eigenvalue}) include both
propagating modes and line waves, numerically we sort them by the proportion
of their norm concentrated in a small region around $x = 0$, neglecting all
but the most confined modes.  Fig. \ref{fig:dispersion}a--b shows the
line wave dispersion relation and field profiles calculated using this method,
which---as shown---agrees with the same calculation carried out using
commercial finite element software (COMSOL multiphysics~\cite{comsol}).

\subsection{Approximations to the integral kernel}

The numerical results shown in Fig. \ref{fig:dispersion} show that the line
wave has some superficial similarities with an electromagnetic surface
wave.  For the surface plasmon/magnon, the asymptotic limit, $k_z / k_0
\rightarrow \infty$ requires two bulk media with zero average
permittivity/permeability.  Equivalently Fig. \ref{fig:dispersion}a shows
that for line waves, the average surface reactance tends to zero in the same
limit.  This similarity was also discussed in the asymptotic analysis of
Ref.~\cite{horsley2014}.  Moreover the field profiles shown in Fig.
\ref{fig:dispersion}b suggest that we might make an exponential approximation
to the form of the line wave, equivalent to the exponential localization of a
surface plasmon/magnon around the interface. In this section we show that we can develop local approximations to the integral kernel $K (x - x')$.  The leading order approximation then yields a line mode that is the exact two dimensional
equivalent of the surface plasmon/magnon, with a dispersion relation that
closely matches the numerical solution to Eq. (\ref{eq:fourier-eigenvalue}).

\begin{figure}[h]
  \raisebox{0.0\height}{\includegraphics[width=15.9966548602912cm,height=5.33221828676374cm]{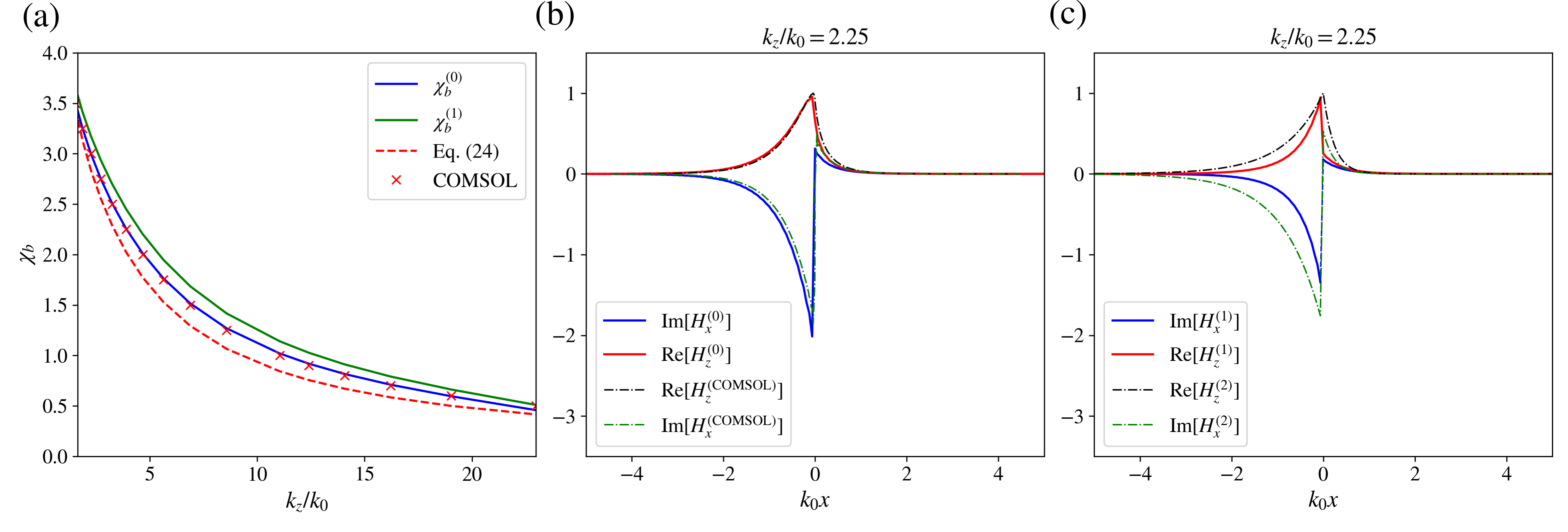}}
  \caption{Description as in Fig. \ref{fig:dispersion}, but for the increased
  reactance contrast $\Delta \chi = 8.309$, illustrating that the approximate
  kernels (\ref{eq:K1}--\ref{eq:K2}) give reduced accuracy for larger contrast
  reactance surfaces.\label{fig:dispersion-2}}
\end{figure}

For inhomogeneous media the difficulty in developing an algebraically simple solution to the non--local equation (\ref{eq:eigenvalue-problem}) can be traced back to the square root denominator in the Fourier representation of the integral kernel (\ref{eq:kernel-defn}).  Were the square root denominator replaced with a polynomial $a_n k^n + a_{n - 1} k^{n - 1} + \cdots$, the inverse of the integral operator would be a differential operator $a_n \left( - \text{i}
\partial_x \right)^n + a_{n - 1} \left( - \text{i} \partial_x \right)^{n - 1}
+ \ldots$, meaning that we could re--write (\ref{eq:eigenvalue-problem}) as a
differential equation.  However, the branch cuts in the square root
denominator make this simplification impossible.  To make progress we
recognise that the Fourier transform of an exponentially localized field (such
as the line waves we are seeking to describe) is centred around zero
wave--vector.  We therefore replace the square root denominator in the
Fourier representation of the kernel (\ref{eq:kernel-defn}) with its series
expansion.  Here we consider two such approximations,
\begin{equation}
  \sqrt{\xi^2 + 1} \sim \left\{ \begin{array}{lc}
    1 + \frac{1}{2} \xi^2 & (\tmop{Approx} . 1)\\
    & \\
    1 & (\tmop{Approx} . 2) .
  \end{array} \right. \label{eq:square-root-series}
\end{equation}
These approximations transform the integrand in (\ref{eq:kernel-defn}), which
contains two square root branch cuts at $\xi = \pm \text{i}$ into a form that
is either an entire function of $k$, or contains simple poles.\quad The
corresponding approximate expressions for the integral kernel
(\ref{eq:kernel-defn}) are then given by
\begin{equation}
  K^{(1)} (x - x') = \int_{- \infty}^{\infty} \frac{\text{d} \xi}{2 \pi}
  \frac{\text{e}^{\text{i} \xi \kappa (x - x')}}{\left( 1 + \frac{1}{2} \xi^2
  \right)} = \frac{\text{e}^{- \sqrt{2} \kappa | x - x' |}}{\sqrt{2}}
  \label{eq:K1} \qquad (\tmop{Approx} . 1)
\end{equation}
and
\begin{equation}
  K^{(2)} (x - x') = \int_{- \infty}^{\infty} \frac{\text{d} \xi}{2 \pi}
  \text{e}^{\text{i} \xi \kappa (x - x')} = \frac{1}{\kappa} \delta (x - x')
  \label{eq:K2} \qquad (\tmop{Approx} . 2)
\end{equation}
As shown in Fig. \ref{fig:kernel}, the first approximation $K^{(1)}$ closely
matches the exact expression (\ref{eq:kernel-defn}), becoming ever more
accurate as the value of $k_z / k_0$ is increased.  The second
approximation (\ref{eq:K2}) only corresponds to the exact kernel in that it
matches the area underneath its curve.  The dispersion relations and field
profiles obtained using these two approximations are compared to the results
of the exact equation (\ref{eq:fourier-eigenvalue}) in Figs.
\ref{fig:dispersion} and \ref{fig:dispersion-2}, illustrating that both
approximations provide a good estimate of the dispersion of line waves.  As shown there, approximation 1 yields fields that have a similar dependence
to the exact solution away from the interface, but all field components are
discontinuous across $x = 0$.  Meanwhile approximation 2 tends to
underestimate the localization of the field.

\subsection{Local line wave equations}\label{sec:local}

The advantage of the approximations (\ref{eq:K1}--\ref{eq:K2}) is that---as
discussed above---they reduce the non--local equation
(\ref{eq:eigenvalue-problem}) to an ordinary differential equation on the
surface: an equation written entirely in terms of the surface fields and their
in--plane derivatives.  For the first approximation (\ref{eq:K1}) we note
that $K^{(1)}$ obeys the inhomogeneous equation $(\partial_x^2 - 2 \kappa^2)
K^{^{(1)}} = - 2 \kappa \delta (x - x')$.  Therefore, applying the
differential operator $\partial_x^2 - 2 \kappa^2$ to Eq.
(\ref{eq:eigenvalue-problem}) yields a local surface wave equation holding on
the $y = 0$ surface
\begin{equation}
    (\tmmathbf{\nabla}_{| |} \times \tmmathbf{\nabla}_{| |} \times - k_0^2) 
    \tmmathbf{H}_{| |} (x) - \frac{k_0}{2 \kappa} \left(
    \frac{\text{d}^2}{\text{d} x^2} - 2 \kappa^2 \right) \chi (x) 
    \tmmathbf{H}_{| |} = 0 \qquad (\tmop{Approx} . 1)\label{eq:approx-1}
\end{equation}
Similarly, using the more extreme approximation (\ref{eq:K2}) where the
integral kernel is proportional to a delta function, Eq.
(\ref{eq:eigenvalue-problem}) becomes
\begin{equation}
  \left[ \tmmathbf{\nabla}_{| |} \times \tmmathbf{\nabla}_{| |} \times -
  k_0^2 \left( 1 - \frac{\kappa}{k_0} \chi (x) \right) \right] \cdot
  \tmmathbf{H}_{| |} (x) = 0. \qquad (\tmop{Approx} . 2) \label{eq:approx-2}
\end{equation}
Note that this equation is equivalent to our earlier one for propagation on a
homogeneous surface (\ref{eq:uniform-surface}) but with $k = 0$, and the
reactance promoted to a position dependent quantity.  It is also worth
noticing that taking the $k_z / k_0 \rightarrow \infty$ limit of the first
approximate equation (\ref{eq:approx-1}) yields the second approximation
(\ref{eq:approx-2}).

\subsection{Fields and dispersion relations}

From the above two surface wave equations (\ref{eq:approx-1}--\ref{eq:approx-2}) we can find analytic expressions for the field profiles and dispersion relations of line waves that closely match the results of finite element simulations.  In this approximation the field exponentially decays away from $x = 0$, and line waves are the direct one--dimensional analogues of surface waves.

We concentrate on the simplest case, Eq. (\ref{eq:approx-2}), although exactly
the same argument can be carried out for the more accurate equation
(\ref{eq:approx-1}) (see Appendix B).  Taking the $x$ component of Eq.
(\ref{eq:approx-2}) we find that the $x$ component of the magnetic field is
determined by the derivative of $H_z$
\begin{equation}
  H_x = - \frac{ \text{i} k_z}{\kappa^2 + k_0 \kappa \chi (x)} \frac{\text{d}
  H_z}{\text{d} x} \label{eq:Hx-condition} .
\end{equation}
This expression for $H_x$ allows us to write the approximate surface wave
equation (\ref{eq:approx-2}) in terms of $H_z$ alone
\begin{equation}
  \frac{\text{d} }{\text{d} x} \left( \frac{k_0^2 - k_0 \kappa \chi
  (x)}{\kappa^2 + k_0 \kappa \chi (x)} \right) \frac{\text{d} H_z}{\text{d} x}
  - k_0^2 \left( 1 - \frac{\kappa}{k_0} \chi (x) \right) H_z = 0.
  \label{eq:Hz-equation}
\end{equation}
Although we are describing the electromagnetic field on a surface, Eq.
(\ref{eq:Hz-equation}), which governs the cross--sectional form of the line
wave, is equivalent to the familiar Helmholtz equation for propagation along a
single axis in a bulk electromagnetic material where the effective
permittivity is given by
\begin{equation}
  \varepsilon_{\tmop{eff}} = - \frac{\left( \frac{\kappa}{k_0} \right)^2 +
  \frac{\kappa}{k_0} \chi (x)}{1 - \frac{\kappa}{k_0} \chi (x)}
  \label{eq:eps-eff}
\end{equation}
and the effective permeability, $\mu_{\tmop{eff}}$ equals
(\ref{eq:mueff}).  By analogy with bulk electromagnetic waves, we can
predict the presence of line waves.  These waves require provided the product
$\varepsilon_{\tmop{eff}} \mu_{\tmop{eff}}$ is negative on both sides of the
interface (for exponential decay, rather than propagation), in addition to
$\varepsilon_{\tmop{eff}}$ changing sign across the interface (ensuring the
decay constant changes sign across $x = 0$).

Although the above derivation holds for any spatially varying reactance $\chi
(x)$, we consider the special case of the step change in impedance given by
(\ref{eq:reactance-line-wave}).  From a cursory inspection of Eq.
(\ref{eq:Hz-equation}) we see that in order that $H_z$ is a well defined
solution, both $H_z$ and $\varepsilon_{\tmop{eff}}^{- 1} \partial_x H_z$ must
be continuous across the impedance interface at $x = 0$.  Given that the
impedance is homogeneous everywhere except at $x = 0$, we can thus write the
solution to Eq. (\ref{eq:Hz-equation}) as a piece--wise function
\begin{equation}
  H_z = H_0 \begin{cases}
    \text{e}^{\beta_L x} & \qquad (x < 0)\\
    & \\
    \text{e}^{- \beta_R x} & \qquad (x > 0)
  \end{cases} \label{eq:line-wave-Hz}
\end{equation}
where, from substitution into (\ref{eq:Hz-equation}) we determine the decay
constants to be given by \ $\beta_{L, R} = \sqrt{\kappa^2 + k_0 \kappa
\chi^{(L, R)}}$.\quad In this approximation the existence of the line wave
thus requires the positivity of $\kappa^2 + k_0 \kappa \chi^{(L, R)}$ on
either side of the $x = 0$ interface.\quad Applying the second condition, for
continuity of $\varepsilon_{\tmop{eff}}^{- 1} \partial_x H_z$ then yields the
line--wave dispersion relation
\begin{equation}
  \left( 1 - \frac{\kappa}{k_0} \chi^{(L)} \right) \sqrt{\frac{\kappa}{k_0} +
  \chi^{(R)}} + \left( 1 - \frac{\kappa}{k_0} \chi^{(R)} \right)
  \sqrt{\frac{\kappa}{k_0} + \chi^{(L)}} = 0 \label{eq:analytic-dispersion}
\end{equation}
which---as shown in Figs. \ref{fig:dispersion} and
\ref{fig:dispersion-2}---provides a good estimate of the numerically
determined dispersion relation calculated from the exact Eq.
(\ref{eq:fourier-eigenvalue}), and using finite element simulations.  From
(\ref{eq:analytic-dispersion}) we can see that, as $k_z \rightarrow \infty$,
the dispersion relation reduces to
\begin{equation}
  k_z \rightarrow \infty : \qquad \chi^{(L)} + \chi^{(R)} = 0
  \label{eq:asymptotic-limit-isotropic}
\end{equation}
in agreement with the asymptotic analysis given in~\cite{horsley2014}.  
For complex reactances, this condition implies a PT--symmetric surface with
balanced loss and gain, as found in~\cite{moccia2020}.  We can therefore
see that, despite the apparently complicated mathematical properties of line
waves, they can be reasonably approximated as exponentially decaying fields on
the surface, obeying the two--dimensional equation (\ref{eq:approx-2}) which
could have been anticipated from boldly promoting the dispersion relation on a homogeneous
surface to a wave equation (\ref{eq:uniform-surface}).

\begin{figure}[h]
  \raisebox{0.0\height}{\includegraphics[width=15.9966548602912cm,height=5.33221828676374cm]{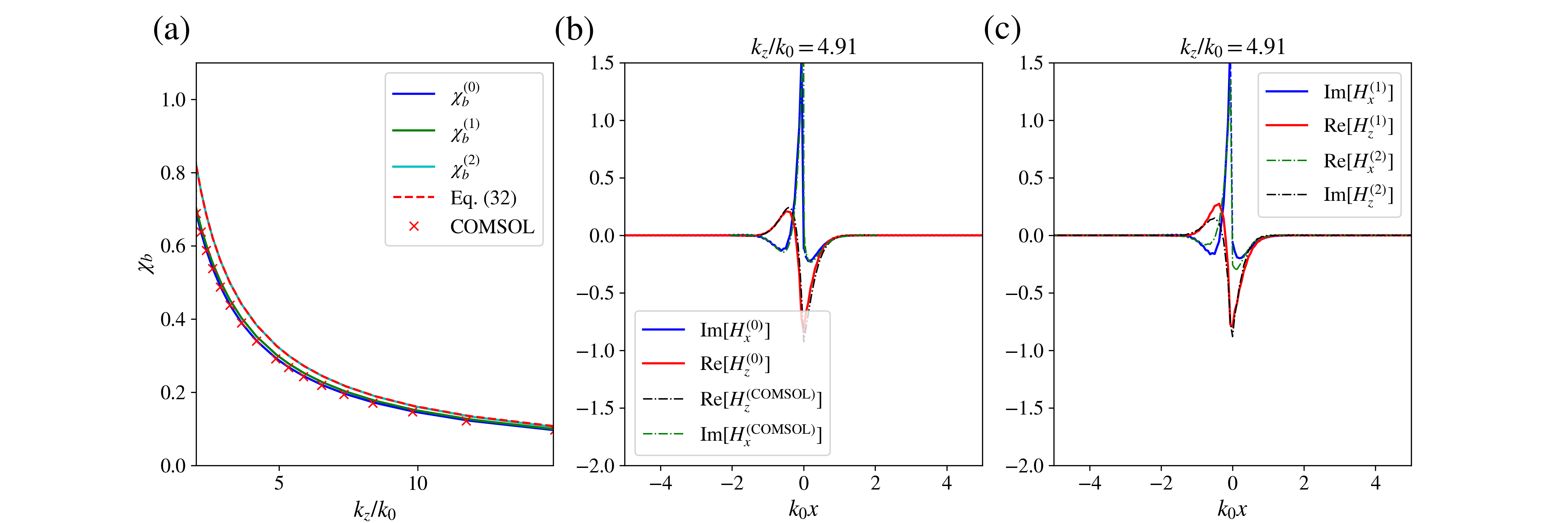}}
  \caption{Line wave profiles on an anisotropic surface with $\Delta
  \tmmathbf{\chi} = \left( \begin{array}{cc}
    2.5 & 1.2\\
    1.2 & 2.5
  \end{array} \right)$ and $\tmmathbf{\chi}_a = 0$, as defined in
  (\ref{eq:anisotropic-eigenvalue}). (a) Dispersion relations obtained using
  the exact integral equation (\ref{eq:anisotropic-eigenvalue}), the two
  approximations (\ref{eq:K1}) and (\ref{eq:K2}), the analytic dispersion
  relation (\ref{eq:anisotropic-dispersion}), and finite element simulations
  (see Appendix A). \ Panels (b) and (c) show that the mode oscillates as it
  decays, characteristic of `ghost' line waves {\cite{moccia2023}}, an effect
  that is still captured by the most extreme approximation to the integral
  kernel (\ref{eq:line-wave-anisotropic}) \label{fig:anisotropic}.}
\end{figure}

\section{Anisotropic surfaces: Effective Gauge fields, and ghost waves}

Finally, having derived this simple theory for isotropic impedance boundaries
we can explore some extensions to more exotic surfaces.  The most obvious
generalization is to anisotropic impedance boundaries.  These were very
recently been discussed in Ref.~\cite{moccia2023}, where primarily using
finite element simulations, one--dimensional ``ghost line waves'' were
discovered: confined waves on a lossless surface that oscillate as they
decay.  Here we shall show that this hybrid behaviour can be
straightforwardly explained using the local approximation to our integral
equation described above.  Put simply, the anisotropy acts as an effective
gauge field on the surface, and just as for an electron subject to a magnetic
vector potential {\cite{peshkin2014}}, this induces extra oscillations in the
surface field.

To treat anisotropic impedance boundaries, we assume the anisotropic
generalization of the surface reactance profile given in Eq.
(\ref{eq:reactance-line-wave})
\begin{equation}
  \tmmathbf{\chi} (x) = \chi_b \tmmathbf{1}_2 + \tmmathbf{\chi}_a +
  \frac{\Delta \tmmathbf{\chi}}{2} \tmop{sign} (x)
  \label{eq:anisotropic-impedance}
\end{equation}
where $\Delta \tmmathbf{\chi}$ is the tensorial difference in the reactance
between the two surfaces, a Hermitian matrix in the case of lossless
surfaces. 6 The matrix $\tmmathbf{\chi}_a$ is a constant matrix
representing the zero--trace part of the average reactance of the two
surfaces, whereas $\chi_b$ is an overall shift to the diagonal elements of the
surface reactance required to support the surface wave.  With this
reactance profile the integral equation (\ref{eq:fourier-eigenvalue}) is
simply generalized to,
\begin{equation}
  \left( \frac{1}{k_0} \tmmathbf{} \frac{\tmmathbf{k} \times \tmmathbf{k}
  \times + k_0^2}{\sqrt{k^2 + \kappa^2}} - \tmmathbf{\chi}_a \right) \cdot
  \widetilde{\tmmathbf{H}}_{| |} (k) + \frac{\Delta \tmmathbf{\chi} \cdot}{2} 
  \frac{\text{i}}{\pi} \text{P}  \int_{- \infty}^{\infty}
  \frac{\widetilde{\tmmathbf{H}}_{| |} (k')}{k - k'} \text{d} k' \cdot
  \widetilde{\tmmathbf{H}}_{| |} (k) = \chi_b^{(0)} 
  \widetilde{\tmmathbf{H}}_{| |} (k) . \label{eq:anisotropic-eigenvalue}
\end{equation}
This remains a Hermitian operator for Hermitian $\Delta \tmmathbf{\chi}$ and
$\tmmathbf{\chi}_a$, indicating eigenmodes are still supported for real values
of the diagonal reactance components, $\chi_b^{(0)}$.

By an identical argument to that presented in Sec.
\ref{sec:local}, the more extreme of the two local approximations to this
integral equation is also a straightforward generalization of our surface
vector Helmholtz equation (\ref{eq:approx-2}), where the reactance is replaced
with a $2 \times 2$ matrix,
\begin{align}
    & (\tmmathbf{\nabla}_{| |} \times \tmmathbf{\nabla}_{| |} \times - k_0^2)
    \tmmathbf{H}_{| |} (x) + \kappa k_0  \tmmathbf{\chi} \cdot
    \tmmathbf{H}_{| |} = 0\nonumber\\
    \rightarrow & H_x = - \frac{1}{\kappa^2 + k_0 \kappa \chi_{x x}}  \left( i
    k_z \frac{\text{d} }{\text{d} x} + k_0 \kappa \chi_{x z} \right) H_z .
    \label{eq:Hx-anisotropic}
\end{align}
The second line follows from taking the $x$--component of the Helmholtz
equation, which determines the component of the magnetic field orthogonal to
the propagation axis.  Interestingly, a comparison with our earlier Eq.
(\ref{eq:Hx-condition}) shows that the off--diagonal element $\chi_{x z}$
induces an effective gauge potential in (\ref{eq:Hx-anisotropic}), modifying
the derivative to $\partial_x - \text{i} k_z^{- 1} k_0 \kappa \chi_{x
z}$.  The appearance of an effective gauge field in the line-wave equation
due to the anisotropy is confirmed after substituting Eq.
(\ref{eq:Hx-anisotropic}) into the preceding vector Helmholtz equation, where
we are left with a second order equation for the field component $H_z$
\begin{equation}
  \left( \frac{d}{d x} - \text{i} A \right) \frac{1}{\varepsilon_{\tmop{eff}}}
  \left( \frac{d}{d x} - \text{i} A \right) H_z + k_0^2 \mu_{\tmop{eff}} H_z =
  0 \label{eq:line-wave-anisotropic} .
\end{equation}
Here we have assumed a symmetric reactance $\chi_{x z} = \chi_{z x}$ which is
equivalent to assuming time reversal symmetry of the surface (for a Hermitian
reactance, both $A$ and its complex conjugate appear in \
(\ref{eq:line-wave-anisotropic})).  In this case the effective gauge
potential is given by $A = \sigma k_z \kappa k_0^{-1} \chi_{x z}$, where $\sigma =
[1 -  (\kappa/k_0) \chi_{x x}]^{- 1}$.\quad Meanwhile the effective
permittivity and permeability are generalized from the earlier expressions
(\ref{eq:mueff}) and (\ref{eq:eps-eff}) to $\varepsilon_{\tmop{eff}} = -
\sigma [(\kappa / k_0)^2 + (\kappa / k_0) \chi_{x x}]$, and $\mu_{\tmop{eff}}
= \sigma [1 - (\kappa / k_0) \tmop{Tr} (\tmmathbf{\chi}) + (\kappa / k_0)^2
\det (\tmmathbf{\chi})]$.

The Helmholtz equation (\ref{eq:line-wave-anisotropic}) takes the form of the
one dimensional Schr{\"o}dinger equation for a charged particle in a magnetic
gauge potential $A$ with a mass proportional to $\varepsilon_{\tmop{eff}}$ and
a scalar potential proportional to $\mu_{\tmop{eff}}$~\cite{peshkin2014}.  The effect of the anisotropy on the surface field
thus both modifies the dispersion relation (\ref{eq:analytic-dispersion}) due
to the generalized forms of $\varepsilon_{\tmop{eff}}$ and $\mu_{\tmop{eff}}$,
as well as introducing field oscillations along the $x$--axis.  These
oscillations due to the gauge field can be isolated through making the
substitution, $H_z = \exp \left( \text{i} \int_0^x A  (x') \text{d} x' \right)
h_z$. 6 This substitution transforms the field equation
(\ref{eq:line-wave-anisotropic}), to an equation for $h_z$ that takes the
earlier form (\ref{eq:Hz-equation})
\begin{equation}
  \frac{d}{d x} \frac{1}{\varepsilon_{\tmop{eff}}} \frac{d h_z}{d x} + k_0^2
  \mu_{\tmop{eff}} h_z = 0 \label{eq:line-mode-no-gauge-field} .
\end{equation}
The only difference from our earlier analysis are the modifications to the
functional form of the effective permittivity and permeability.\quad Assuming
$\varepsilon_{\tmop{eff}} \mu_{\tmop{eff}} < 0$, Eq.
(\ref{eq:line-mode-no-gauge-field}) again admits confined line--wave solutions
$h_z$ of the form (\ref{eq:line-wave-Hz}) with decay constants
\begin{align}
    \beta_{L, R} & = k_0 \sqrt{- \varepsilon_{\tmop{eff}} \mu_{\tmop{eff}}}\\
    & \nonumber\\
    & = k_0 \frac{\sqrt{\left( \frac{\kappa}{k_0} \right)^2 +
    \frac{\kappa}{k_0} \chi_{x x}^{(L, R)}}}{\left| 1 - \frac{\kappa}{k_0}
    \chi_{x x}^{(L, R)} \right|} \sqrt{1 - \frac{\kappa}{k_0} \tmop{Tr}
    [\tmmathbf{\chi}^{(L, R)}] + \left( \frac{\kappa}{k_0} \right)^2 \det
    [\tmmathbf{\chi}^{(L, R)}]}
\end{align}
which reduce to those given below Eq. (\ref{eq:line-wave-Hz}) for isotropic surfaces where $\chi_{x
x}^{(L, R)} = \chi_{z z}^{(L, R)}$ and $\chi_{x z}^{(L, R)} = \chi_{z x}^{(L,
R)} = 0$.  Again demanding the continuity of $\varepsilon_{\tmop{eff}}^{-
1} \partial_x h_z$ yields the corresponding line wave dispersion relation
\begin{equation}
  \left( 1 + \frac{k_0}{\kappa} \chi_{x x}^{(L)} \right) \left( 1 -
  \frac{\kappa}{k_0} \chi_{x x}^{(R)} \right) \beta_L + \left( 1 +
  \frac{k_0}{\kappa} \chi_{x x}^{(R)} \right) \left( 1 - \frac{\kappa}{k_0}
  \chi_{x x}^{(L)} \right) \beta_R = 0 \label{eq:anisotropic-dispersion} .
\end{equation}
This is plotted as the red dashed line in Fig. \ref{fig:anisotropic}a, where
it is evident that for this range of impedance values, this approximation
closely matches the results of both the exact integral equation
(\ref{eq:anisotropic-eigenvalue}) and finite element simulations.\quad Figures
\ref{fig:anisotropic}b and \ref{fig:anisotropic}c show cross sections of the
line wave field, where---compared to the isotropic results shown in Figs.
\ref{fig:dispersion}--\ref{fig:dispersion-2}---there is the anticipated hybrid of
oscillation and decay away from the $x = 0$ interface on the surface.

As in the isotropic case discussed above and in Ref. {\cite{horsley2014}}, in
the asymptotic limit $k_z \rightarrow \infty$,the dispersion relation
simplifies to a constraint on the material parameters,
\begin{equation}
  \frac{\chi_{x x}^{(R)}}{| \chi_{x x}^{(L)} |} \sqrt{\det
  [\tmmathbf{\chi}^{(L)}]} + \frac{\chi_{x x}^{(L)}}{| \chi_{x x}^{(R)} |}
  \sqrt{\det [\tmmathbf{\chi}^{(R)}]} = 0
  \label{eq:asymptotic-limit-anisotropic}
\end{equation}
which reduces to our previous result (\ref{eq:asymptotic-limit-isotropic})
when the surface is isotropic.\quad Indeed, for anisotropic surfaces with a
reactance matrix of equal determinant $\det [\tmmathbf{\chi}^{(L)}] = \det
[\tmmathbf{\chi}^{(R)}]$, Eq. (\ref{eq:asymptotic-limit-anisotropic}) predicts
that the asymptotic limit occurs when the diagonal elements $\chi_{x x}^{(L,
R)}$ are of equal magnitude and opposite sign.

\section{Summary and Conclusions}

Although wedge plasmons and edge waves are well--known one--dimensional
excitations, these are typically confined to a surface through the effect of
surface curvature {\cite{valle2010}}.  On the other hand, line waves
\cite{sievenpiper2017, horsley2014, moccia2020, moccia2023}
are one--dimensional waves confined to \tmtextit{flat} surfaces by the
distribution of surface impedance.  Line waves are much more difficult to
theoretically analyse than surface waves, and to--date a simple understanding
of these modes has been lacking.

In this work we have provided a new general theory for analysing the behaviour
of line waves on impedance boundaries.  In the exact case, the properties
of line waves can be calculated through determining the eigenfunctions of the
integral equation (\ref{eq:fourier-eigenvalue}), or its generalization to
anisotropic surfaces (\ref{eq:anisotropic-eigenvalue}).  To gain further
understanding of these waves, this integral equation can be approximated as a
differential equation using the expansions of the kernel (\ref{eq:approx-1})
and (\ref{eq:approx-2}), revealing that the cross--section of the line wave
obeys a one--dimensional Helmholtz equation on the surface, with a spatially
dispersive effective permittivity and permeability.  From this we have
shown that an approximate line wave dispersion relation can be derived exactly
as is done for the surface plasmon/magnon, with the result closely matching
the results of finite element simulations (see Fig. \ref{fig:dispersion}),
deviating more when the impedance contrast is increased to larger values (see
Fig. \ref{fig:dispersion-2}).

As we have shown, it is also straightforward to extend this theory to
anisotropic impedance boundaries.  In this case the cross section of the
line wave also (approximately) obeys a one dimensional Helmholtz equation,
with the anisotropy modifying the effective permittivity and permeability
values, ensuring---for example---that the asymptotic limit of the dispersion
relation no longer requires equal and opposite reactance values, but the more
complicated combination of material parameters given in Eq.
(\ref{eq:asymptotic-limit-anisotropic}).  Interestingly, we have also
found that anisotropy of the reactance induces an effective gauge field on the
surface.  As we are dealing with an effective one--dimensional equation,
this gauge field cannot induce e.g. cyclotron orbits of the surface wave, but
instead is equivalent to the gauge transformation given above Eq.
(\ref{eq:line-mode-no-gauge-field}) which leads to oscillations in the field
away from the interface.  We have verified this numerically (see Fig.
\ref{fig:anisotropic}) using both the exact integral equation and finite
element simulations, providing an explanation for the recently identified
`ghost' line waves {\cite{moccia2023}}.

Just as understanding surfaces waves has led to surface wave antennas and the
field of plasmonics, understanding line waves could lead to similar
applications.  They allow---for instance---electromagnetic energy to be
channeled without the use of a waveguide, and as shown above and in
{\cite{moccia2023}} the flow of this near--field energy can be moulded via the
anisotropy of the surface.  The theory presented here may provide a framework for a deeper understanding these waves in a larger parameter space than has been considered to--date.

\section{Acknowledgements}

SARH and AD thank the Royal Society and TATA for financial support
(RPG-2016-186)).\quad SARH thanks Ian Hooper and James Capers for useful
discussions.

\section{Appendix A: Modifications to comsol}

To simulate anisotropic impedance boundaries we modified the equations of the
``impedance boundary'' in COMSOL Multiphysics.\quad This was done for surfaces
with surface normal $\tmmathbf{\hat{y}}$ by replacing the following
expressions
\begin{verbatim}
emw.imp1.Jsx  = ((Zzz*(emw.tEx+emw.Esx)-Zxz*(emw.tEz+emw.Esz))/dZ)/Z0
emw.imp1.Jsy  = 0
emw.imp1.Jsz  = ((Zxx*(emw.tEz+emw.Esz)-Zzx*(emw.tEx+emw.Esx))/dZ)/Z0   
\end{verbatim}
{\noindent}where \tmverbatim{Z0=377}$\Omega$ is the free space impedance,
\tmverbatim{Zxx}, \tmverbatim{Zxz}, \tmverbatim{Zzx}, and \tmverbatim{Zzz} are
the components of the impedance tensor $\tmmathbf{Z} = i \tmmathbf{\chi}$, and
\tmverbatim{dZ} is the determinant of the impedance tensor, all defined as a
set of parameters for each impedance boundary in the COMSOL model.

\section{AppenDIX B: the more accurate solution}

In the main text we gave analytic results for the very simplest approximation
to the integral kernel (\ref{eq:kernel-defn}).  We can also develop more
accurate analytic results for the next order approximation, although the
formulae are more complicated.  Here we give this more accurate solution
for an isotropic impedance boundary.  The approximate surface wave
equation is given by Eq. (\ref{eq:approx-1}) in the main text, which we repeat
here:
\begin{equation}
  (\tmmathbf{\nabla}_{| |} \otimes \tmmathbf{\nabla}_{| |} -
  \tmmathbf{\nabla}_{| |}^2 - k_0^2)  \tmmathbf{H}_{| |} (x) - \frac{k_0}{2
  \kappa} \left( \frac{\text{d}^2}{\text{d} x^2} - 2 \kappa^2 \right) \chi (x)
  \tmmathbf{H}_{| |} = 0.
\end{equation}
Splitting this equation into components we obtain two coupled Helmholtz
equations, one for the $x$--component
\begin{equation}
  \frac{k_0}{2 \kappa} \frac{\text{d}^2}{\text{d} x^2} (\chi H_x) - (\kappa^2
  + k_0 \kappa \chi) H_x - \text{i} k_z \frac{\text{d} H_z}{\text{d} x} = 0,
  \label{eq:a1x}
\end{equation}
and one for the $z$--component,
\begin{equation}
  \frac{\text{d}^2}{\text{d} x^2} \left( 1 + \frac{k_0}{2 \kappa} \chi \right)
  H_z + k_0^2 \left( 1 - \frac{\kappa}{k_0} \chi \right) H_z - \text{i} k_z
  \frac{\text{d} H_x}{\text{d} x} = 0. \label{eq:a1z}
\end{equation}
Assuming a junction between two surfaces as in (\ref{eq:reactance-line-wave})
we can solve (\ref{eq:a1x}--\ref{eq:a1z}) by assuming---in the homogeneous
regions---that the field has the form
\begin{equation}
  \left( \begin{array}{c}
    H_x\\
    H_z
  \end{array} \right) = \left( \begin{array}{c}
    A_x\\
    A_z
  \end{array} \right) \text{e}^{\pm \beta x}
\end{equation}
($\beta > 0$) so that (\ref{eq:a1x}--\ref{eq:a1z}) can be written as a single
matrix equation
\begin{equation}
  \left( \begin{array}{cc}
    \frac{k_0 \chi}{2 \kappa} \beta^2 - (\kappa^2 + k_0 \kappa \chi) & \mp
    \text{i} k_z \beta\\
    \mp \text{i} k_z \beta & \left( 1 + \frac{k_0}{2 \kappa} \chi \right)
    \beta^2 + k_0^2 \left( 1 - \frac{\kappa}{k_0} \chi \right)
  \end{array} \right) \left( \begin{array}{c}
    A_x\\
    A_z
  \end{array} \right) = 0, \label{eq:matrix-field}
\end{equation}
which requires the matrix to have zero determinant.\quad The zero determinant
condition yields a fourth order polynomial in the decay constant $\beta$
\begin{equation}
  \left[ \frac{k_0 \chi}{2 \kappa} \beta^2 - (\kappa^2 + k_0 \kappa \chi)
  \right] \left[ \left( 1 + \frac{k_0}{2 \kappa} \chi \right) \beta^2 + k_0^2
  \left( 1 - \frac{\kappa}{k_0} \chi \right) \right] + k_z^2 \beta^2 = 0.
  \label{eq:beta-fourth}
\end{equation}
As Eq. (\ref{eq:beta-fourth}) is a quadratic equation in $\beta^2$ we have
four values of $\beta$ that form two pairs of solutions with the same
magnitude
\begin{equation}
  \begin{array}{cl}
    \beta_1^2 & = \frac{1}{2 a} \left[ - b + \sqrt{b^2 - 4 a c} \right]\\
    & \\
    \beta_2^2 & = \frac{1}{2 a} \left[ - b - \sqrt{b^2 - 4 a c} \right] .
  \end{array} \label{eq:two-decay-constants}
\end{equation}
To simplify the notation we have introduced three constants,
\begin{equation}
  \begin{array}{cl}
    a & = \frac{k_0 \chi}{2 \kappa} \left( 1 + \frac{k_0}{2 \kappa} \chi
    \right)\\
    & \\
    b & = \frac{k_0 \chi}{2 \kappa} k_0^2 \left( 1 - \frac{\kappa}{k_0} \chi
    \right) - (\kappa^2 + k_0 \kappa \chi) \left( 1 + \frac{k_0}{2 \kappa}
    \chi \right) + k_z^2\\
    & \\
    c & = - (\kappa^2 + k_0 \kappa \chi) k_0^2 \left( 1 - \frac{\kappa}{k_0}
    \chi \right) .
  \end{array}
\end{equation}
Therefore, on each side of the interface our field is composed of a
combination of two exponentials with decay constants $\beta_1$ and $\beta_2$
given by Eq. (\ref{eq:two-decay-constants}).  Each of these decaying
components is multiplied by the corresponding zero eigenvector of the matrix
equation (\ref{eq:matrix-field}) so that on each side of the interface the
surface magnetic field takes the form
\begin{multline}
  x < 0 : \qquad \tmmathbf{H}_{| |} = c_1^{(L)} N_1^{(L)} \left(
  \begin{matrix}
    \text{i} k_z \beta_1^{(L)}\\
    \\
    \frac{k_0 \chi^{(L)}}{2 \kappa} [\beta_1^{(L)}]^2 - (\kappa^2 + k_0 \kappa
    \chi^{(L)})
  \end{matrix} \right) \text{e}^{\beta_1^{(L)} x} \\[10pt]
  + c_2^{(L)} N_2^{(L)} \left(
  \begin{matrix}
    \text{i} k_z \beta_2^{(L)}\\
    \\
    \frac{k_0 \chi^{(L)}}{2 \kappa} [\beta_2^{(L)}]^2 - (\kappa^2 + k_0 \kappa
    \chi^{(L)})
  \end{matrix} \right) \text{e}^{\beta_2^{(L)} x} \label{eq:field-left}
\end{multline}
where $N_{1, 2}^{(L, R)}$ are the normalization constants defined as
\begin{equation}
  N_{1, 2}^{(L, R)} = \frac{1}{\sqrt{\left( \frac{k_0 \chi^{(L, R)}}{2 \kappa}
  [\beta_{1, 2}^{(L, R)}]^2 - (\kappa^2 + k_0 \kappa \chi^{(L, R)}) \right)^2
  + k_z^2 [\beta_{1, 2}^{(L, R)}]^2}}
\end{equation}
The constants $c_{1, 2}^{(L, R)}$ appearing in Eq. (\ref{eq:field-left}) are
yet to be determined and their represent the amplitudes of the two different
exponentially decaying solutions in the region $x < 0$.\quad The field on the
right hand side of the $x = 0$ line is similarly expressed in terms of two
decaying solutions
\begin{multline}
  x > 0 : \qquad \tmmathbf{H}_{| |} = c_1^{(R)} N_1^{(R)} \left(
  \begin{matrix}
    - \text{i} k_z \beta_1^{(R)}\\
    \\
    \frac{k_0 \chi^{(R)}}{2 \kappa} [\beta_1^{(R)}]^2 - (\kappa^2 + k_0 \kappa
    \chi^{(R)})
  \end{matrix} \right) \text{e}^{- \beta_1^{(R)} x}\\[10pt]
  + c_2^{(R)} N_2^{(R)}
  \left( \begin{matrix}
    - \text{i} k_z \beta_2^{(R)}\\
    \\
    \frac{k_0 \chi^{(R)}}{2 \kappa} [\beta_2^{(R)}]^2 - (\kappa^2 + k_0 \kappa
    \chi^{(R)})
  \end{matrix} \right) \text{e}^{- \beta_2^{(R)} x} \label{eq:field-right}
\end{multline}
To determine the four unknown expansion coefficients in Eqns.
(\ref{eq:field-left}--\ref{eq:field-right}), plus the dispersion relation we
must apply four boundary conditions at $x = 0$.  These four boundary
conditions are contained in the two coupled differential equations
(\ref{eq:a1x}) and (\ref{eq:a1z}) derived above.  For the solution to hold
across the line $x = 0$, all differentiated quantities must be
continuous.  Therefore, from examining the second derivatives in Eqns.
(\ref{eq:a1x}) and (\ref{eq:a1z}) we can see that the first derivatives they
contain will only be well defined if $\chi (x) H_x$ and $(1 + k_0 \chi / 2
\kappa) H_z$ are continuous.  Similarly, integrating these equations across
an infinitesimal interval containing $x = 0$ we can also see that $\partial_x
(k_0 \chi H_x / 2 \kappa) - \text{i} k_z H_z$ and $\partial_x (1 + k_0 \chi /
2 \kappa) H_z - \text{i} k_z H_x$ are continuous across the interface.  These are the four conditions we require to determine the dispersion of the
line wave to a more accurate approximation.

\begin{figure}[h]
  \raisebox{0.0\height}{\includegraphics[width=8.01575823166732cm,height=6.02923717696445cm]{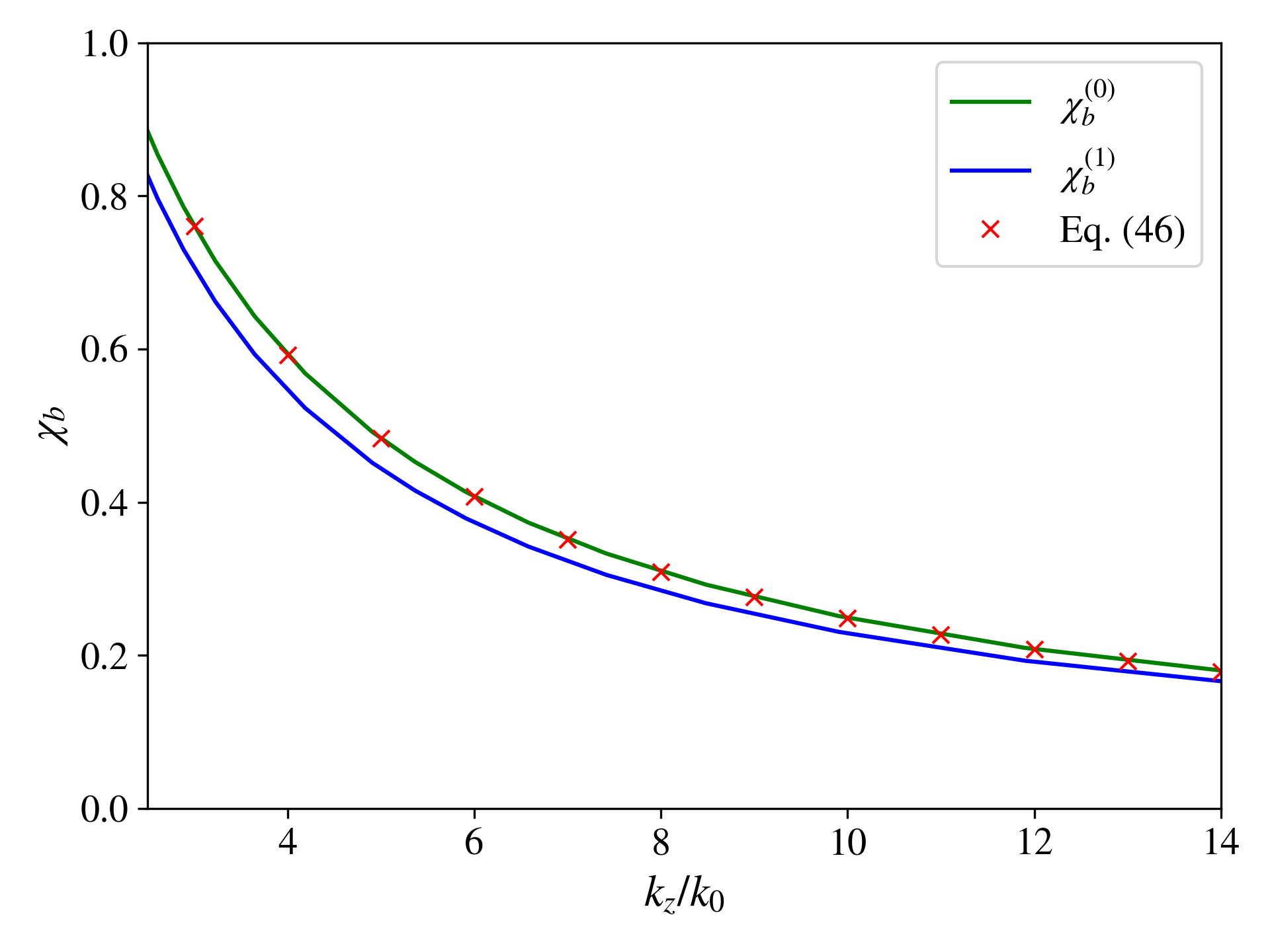}}
  \caption{Line wave dispersion for $\Delta \chi = 3.209$, with other
  parameters as in Fig. \ref{fig:dispersion}.\quad The analytic expression
  (\ref{eq:accurate-dispersion}) reproduces the result of using the
  approximate integral kernel $K^{(1)}$ given in Eq.
  (\ref{eq:K1}).\label{fig:check-a1}}
\end{figure}

Applying these four conditions to the expansions (\ref{eq:field-left}) and
(\ref{eq:field-right}) we find the following $4 \times 4$ matrix
representation of the boundary conditions,
\begin{equation}
\footnotesize
  \left( \begin{matrix}
    N_1^{(L)} \beta_1^{(L)} \chi^{(L)} & N_2^{(L)} \beta_2^{(L)} \chi^{(L)} &
    N_1^{(R)} \beta_1^{(R)} \chi^{(R)} & N_2^{(R)} \beta_2^{(R)} \chi^{(R)}\\
    N_1^{(L)} \Sigma_1^{(L)} & N_2^{(L)} \Sigma_2^{(L)} & - N_1^{(R)}
    \Sigma_1^{(R)} & - N_2^{(R)} \Sigma_2^{(R)}\\
    - N_1^{(L)} (\kappa^2 + k_0 \kappa \chi^{(L)}) & - N_2^{(L)} (\kappa^2 +
    k_0 \kappa \chi^{(L)}) & N_1^{(R)} (\kappa^2 + k_0 \kappa \chi^{(R)}) &
    N_2^{(R)} (\kappa^2 + k_0 \kappa \chi^{(R)})\\
    N_1^{(L)} \beta_1^{(L)} (\Sigma_1^{(L)} + k_z^2) & N_2^{(L)} \beta_2^{(L)}
    (\Sigma_2^{(L)} + k_z^2) & N_1^{(R)} \beta_1^{(R)} (\Sigma_1^{(R)} +
    k_z^2) & N_2^{(R)} \beta_2^{(R)} (\Sigma_2^{(R)} + k_z^2)
  \end{matrix} \right) \left( \begin{matrix}
    c_1^{(L)}\\
    c_2^{(L)}\\
    c_1^{(R)}\\
    c_2^{(R)}
  \end{matrix} \right) = 0
\end{equation}
where $\Sigma_{1, 2}^{(L, R)} = \left( 1 + \frac{k_0 \chi^{(L, R)}}{2 \kappa}
\right) \left[ \frac{k_0 \chi^{(L, R)}}{2 \kappa} [\beta_{1, 2}^{(L, R)}]^2 -
(\kappa^2 + k_0 \kappa \chi^{(L, R)}) \right]$.\quad The vanishing determinant
of this matrix defines the dispersion relation of these modes,
\begin{equation}
\footnotesize
  \det \left( \begin{matrix}
    \beta_1^{(L)} N_1^{(L)} \chi^{(L)} & \beta_2^{(L)} N_2^{(L)} \chi^{(L)} &
    \beta_1^{(R)} N_1^{(R)} \chi^{(R)} & \beta_2^{(R)} N_2^{(R)} \chi^{(R)}\\
    N_1^{(L)} \Sigma_1^{(L)} & N_2^{(L)} \Sigma_2^{(L)} & - N_1^{(R)}
    \Sigma_1^{(R)} & - N_2^{(R)} \Sigma_2^{(R)}\\
    - N_1^{(L)} (\kappa^2 + k_0 \kappa \chi^{(L)}) & - N_2^{(L)} (\kappa^2 +
    k_0 \kappa \chi^{(L)}) & N_1^{(R)} (\kappa^2 + k_0 \kappa \chi^{(R)}) &
    N_2^{(R)} (\kappa^2 + k_0 \kappa \chi^{(R)})\\
    \beta_1^{(L)} N_1^{(L)} (\Sigma_1^{(L)} + k_z^2) & \beta_2^{(L)} N_2^{(L)}
    (\Sigma_2^{(L)} + k_z^2) & \beta_1^{(R)} N_1^{(R)} (\Sigma_1^{(R)} +
    k_z^2) & \beta_2^{(R)} N_2^{(R)} (\Sigma_2^{(R)} + k_z^2)
  \end{matrix} \right) = 0, \label{eq:accurate-dispersion}
\end{equation}
which is the analytic expression for the more accurate dispersion relation
shown in Figs. \ref{fig:dispersion}--\ref{fig:dispersion-2}.\quad Fig.
\ref{fig:check-a1} shows---for an arbitrarily chosen value of $\Delta
\chi$---that this expression reproduces the results of the integral equation
(\ref{eq:eigenvalue-problem}) with the approximate kernel (\ref{eq:K1}).

\bibliographystyle{plain}
\bibliography{refs} 

\end{document}